\documentclass[aps,twocolumn,showpacs]{revtex4}
\usepackage{bm}
\usepackage{graphicx}
\usepackage{epsfig}
\begin{document}

\title{Direct current driven by ac electric field in quantum wells}
\author{S.\,A.\,Tarasenko}
\affiliation{A.F.~Ioffe Physical-Technical Institute, Russian Academy of Sciences, 194021 St.~Petersburg, Russia}




\begin{abstract}

It is shown that the excitation of charge carriers by ac electric field with zero average driving leads to a direct electric current in quantum well structures. The current emerges for both linear and circular polarization of the ac electric field and depends on the field polarization and frequency. We present a micoscopic model and an analytical theory of such a nonlinear electron transport in quantum wells with structure inversion asymmetry. In such systems, dc current is induced by ac electric field which has both the in-plane and out-of-plane components. The ac field polarized in the interface plane gives rise to a direct current if the quantum well is subjected to an in-plane static magnetic field.  
 
\end{abstract}

\maketitle

\section{Introduction}

The excitation of charge carriers by ac electric field in noncentrosymmetric semiconductor structures may lead to a direct electric current even in the absence of dc driving.
In the high-frequency spectral range, when mechanisms of the current formation involve quantum
optical transitions, such effects are usually referred to as photogalvanic effects~\cite{Sturman_book,IvchenkoGanichev}. At present, they are intensively studied in various low-dimensional systems and provide the insight into
the band structure details as well as the kinetics of photoexcited carriers~\cite{Giglberger07,Olbrich09,Zhang09,Dai10,Tarasenko07,Takhtamirov09}.
In the classical frequency range, i.e., $\omega \ll \bar{\varepsilon}/\hbar$, where $\omega$ is the field frequency and $\bar{\varepsilon}$ is the mean kinetic energy of carriers,
the photogalvanic effects can be fruitfully treated as nonlinear electron transport and included in the more general class of quantum ratchets (for a review see~\cite{Reimann02}). The ratchet transport of charge carriers induced by ac electric field 
was studied both theoretically and experimentally for semiconductor structures with artificially fabricated asymmetric scatterers~\cite{Entin06,Sassine08}. Recently, it has also been addressed theoretically for bulk wurtzite crystals and low-symmetry quantum wells (QWs) based on zinc-blende-lattice compounds~\cite{Deyo09,Moore10}. However, the proposed mechanisms of the current formation require the multiband mixing of states  
and, therefore, vanish in the effective-mass approximation.

Here, we show that in conventional quantum wells, where the space inversion is lifted
by structure asymmetry, the ratchet transport of free carriers emerges in the simple one-band model of size-quantized states. We develop a microscopic model and an analytical theory of the direct current generation by ac electric field, which are valid for the classical range of the field frequency. It is shown that the electric current emerges for both linear and circular polarization of the ac field and its magnitude depends on the radiation frequency. We also study the effect of an in-plane static magnetic field on the electron transport and show that the magnetic field gives rise to additional contributions to the dc current which have different polarization dependences. For simplicity, we neglect spin splitting of the conduction band and focus on the orbital mechanisms of the current formation. 

Phenomenologically, the density of direct current $\bm{j}$ induced by ac electric field 
is described by
\begin{equation}\label{j_phen}
j_{\alpha} = \sum_{\beta\gamma} \chi_{\alpha\beta\gamma} E_{\beta} E_{\gamma}^* + \sum_{\beta\gamma\delta} \phi_{\alpha\beta\gamma\delta} B_{\beta} E_{\gamma} E_{\delta}^* \:,
\end{equation} 
where $\bm{E}$ is the complex amplitude of the electric field
\begin{equation}\label{E_time}
\bm{E}(t)=\bm{E} \exp(-i\omega t) + \bm{E}^* \exp(+i\omega t) 
\end{equation} 
assumed to be homogeneous, the indices $\alpha$, $\beta$, $\gamma$, and $\delta$ enumerate the Cartesian coordinates, and components of the tensors $\chi$ and $\phi$ satisfy the relations $\chi_{\alpha\beta\gamma}=\chi_{\alpha\gamma\beta}^*$ and $\phi_{\alpha\beta\gamma\delta}=\phi_{\alpha\beta\delta\gamma}^*$, respectively, which follow from the reality of the current density $\bm{j}$. The third-rank tensor $\chi$ describes the photogalvanic effect (or high-frequency nonlinear conductivity). The model and miscroscopic theory of this effect are presented in Sec.~\ref{PGE}. The forth-rank tensor $\phi$ is responsible for additional current contributions emerging in the presence of an external static magnetic field $\bm{B}$; they are addressed in Sec.~\ref{MPGE}.

\section{High-frequency nonlinear conductivity}\label{PGE}

We consider a semiconductor quantum well with structure inversion asymmetry and assume that the well is isotropic in the interface plane ($C_{\infty v}$ point-group symmetry).   
Straightforward symmetry analysis yields that the in-plane dc current in such structures can be induced only by ac field which has both the in-plane $\bm{E}_{\parallel}=(E_x,E_y)$ and out-of-plane $E_z$ components. The tensor $\chi$ has nonzero components  $\chi_{xxz}=\chi_{yyz}=\chi_{xzx}^*=\chi_{yzx}^*$, and the current density given by the first term in the right-hand side of Eq.~(\ref{j_phen}) can be rewritten in the form
\begin{equation}\label{j_PGE_phen}
\bm{j} =  L \,(\bm{E}_{\parallel} E_z^* + E_z \bm{E}_{\parallel}^*) + C \,i (\bm{E}_{\parallel} E_z^* - E_z \bm{E}_{\parallel}^*) \:.
\end{equation} 
Here, the phenomenological parameter $L={\rm Re} \chi_{xxz}$ describes the electric current that is induced by linearly polarized ac field and insensitive to the sign of radiation helicity for elliptical polarization. In contrast, $C={\rm Im} \chi_{xxz}$ stands for the radiation-helicity dependent electric current  vanishing for the linearly polarized radiation. This is due to the fact that $i (E_{\alpha} E_{\beta}^* - E_{\beta} E_{\alpha}^*)$ ($\alpha \neq \beta$) is nothing but a component of the pseudovector $i[\bm{E}\times \bm{E}^*]=|\bm{E}|^2 (\bm{q}/q) P_{circ}$, where $\bm{q}$ and $P_{circ}$ are the wave vector and the circular polarization degree of the electromagnetic wave. 

\begin{figure}[b]
 \centering\includegraphics[width=0.45\textwidth]{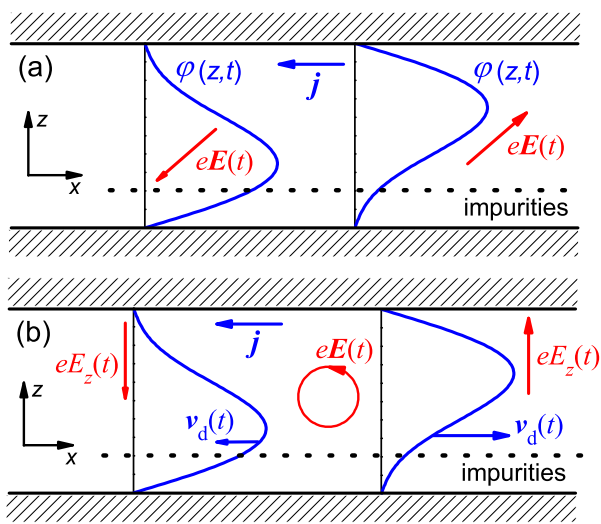}
 \caption{(Color online) Microscopic model of dc current $\bm{j}$ generation by (a) linearly polarized and (b) circularly polarized ac electric field $\bm{E}(t)$ in a quantum well.}
 \label{figure1}
\end{figure}

The microscopic mechanisms of the current generation are illustrated in Fig.~\ref{figure1}a and~\ref{figure1}b for linearly and circularly polarized ac field, respectively. We assume that the electric field $\bm{E}(t)$ is polarized in the $(x,z)$ plane and the electron mobility is limited by electron scattering from static impurities. The structure inversion asymmetry is modeled here by placing the $\delta$-layer of impurities (dotted line) closer to the lower interface rather than exactly in the QW center. However, the microscopic model presented below is also valid for QWs where nonequivalence of $z$ and $-z$ directions is achieved by the asymmetry of confinement potential.

Figure~\ref{figure1}a sketches the mechanism of dc current generation by linearly polarized ac field, i.e., when the in-plane $E_x(t)$ and out-of-plane $E_z(t)$ components are co-phased.
The in-plane oscillating field $E_x(t)$ causes an alternating 
current of electrons along the $x$ axis.
The time-average value of the force $e E_x(t)$ acting upon carriers is zero, therefore, dc current driven solely by the in-plane electric field would vanish.
However, the electric field $\bm{E}(t)$ has also the out-of-plane component $E_z(t)$ which oscillates at the same frequency and acts upon the charge carriers as well. The force $e E_z(t)$ pushes the carriers to the upper or lower interface depending on the force direction and thereby changes the electron wave function along the QW normal $\varphi(z,t)$, see Fig.~\ref{figure1}a. Such time dependence of the envelope function $\varphi(z,t)$ results, in turn, in the modulation of the electron mobility $\mu(t)$ at the frequency $\omega$ because the impurities, which determine the mobility, are shifted off the QW center in our model. 
Therefore, when the carriers are driven in one direction in the QW plane 
their mobility and, hence, the drift velocity $\bm{v}_d$ are higher than those a half period later when the carriers flow in the opposite direction.
Such an asymmetry in the drift of charge carriers along $x$ and $-x$ directions implies a non-vanishing dc electric current.

For circularly polarized ac electric field (Fig.~\ref{figure1}b), the in-plane and out-of-plane components of $\bm{E}(t)$ are phase-shifted by $\pm\pi/2$:
the component $E_z(t)$ reaches maximum when $E_x(t)$ is zero and vice versa. Now, the time-average product $E_x(t) E_z(t)$ vanishes and no dc current emerges in the static limit $\omega \rightarrow 0$. The time-average flow of carriers along the $x$ axis is obtained only at finite frequency if one takes into account the retardation of the drift velocity $\bm{v}_d(t)$ with respect to the in-plane field $E_x(t)$. Indeed, as is well known from the classical Drude theory of high-frequency conductivity, $\bm{v}_d(t)$ does not follow  $\bm{E}_{\parallel}(t)$ exactly but is behind the field with the retardation phase shift of $\arctan (\omega\tau_p)$, where $\tau_p$ is the momentum relaxation time. Due to the retardation, the carriers keep moving in the QW plane even 
when $E_x(t)=0$ and the field component $E_z(t)$ efficiently affects the mobility, see Fig.~\ref{figure1}b.
Similarly to the case of linearly polarized field, such modulation of the mobility leads to a time-average drift current along the $x$ axis. An interesting feature of dc current induced by circularly polarized ac field is that the current direction is opposite for right-handed ($\sigma^+$) and left-handed ($\sigma^-$) polarization. Indeed, for $\sigma^+$ and $\sigma^-$ polarizations the phase shift between the field components $E_x(t)$ and $E_z(t)$ has opposite sign. Therefore, the inversion of radiation helicity changes the sign of the mobility oscillations and reverses the electric current.   

The expression for dc electric current can be readily derived in the framework of the classical Drude theory. In this approach, time evolution of the in-plane drift velocity $\bm{v}_d(t)$ is found from the Newton equation
\begin{equation}\label{v_drift}
\frac{d \bm{v}_d(t)}{d t} = \frac{e \bm{E}_{\parallel}(t)}{m^*}  -  \bm{v}_d(t) \gamma(t) \:,
\end{equation}
where $m^*$ is the effective mass and $\gamma(t)$ is the rate of velocity relaxation which depends on time due to the effect of out-of-plane field component $E_z(t)$ on the function of size quantization, see Fig.~\ref{figure1}. In the linear in $E_z(t)$ regime, the time dependence of $\gamma(t)$ follows $E_z(t)$ and can be presented in the form
\begin{equation}\label{gamma}
\gamma(t) = 1/\tau_p + \zeta e E_z(t) \:,
\end{equation}
where $\tau_p$ is the velocity (momentum) relaxation time at $E_z=0$ and $\zeta$ is a constant to be calculated below. Note, that both $\bm{E}_{\parallel}(t)$ and $E_z(t)$ oscillate at the same frequency [see Eq.~(\ref{E_time})], therefore the drift velocity $\bm{v}_d(t)$ contains harmonics at zero as well as the double frequencies. To solve Eq.~(\ref{v_drift}) we decompose the drift velocity into harmonics
\begin{equation}
\bm{v}_d(t) = \sum\limits_{n=0,\pm1,\ldots} \bm{v}_d^{(n)} \exp(- i n \omega t) \:
\end{equation}
and finally obtain within linear in $E_z$ and $\bm{E}_{\parallel}$ approximation
\begin{equation}\label{v_0}
\bm{v}_d^{(0)} = - \zeta e \tau_p \left( \bm{v}_d^{(1)} E_z^* + \bm{v}_d^{(-1)} E_z \right) \:,
\end{equation}
\begin{equation}\label{v_1}
\bm{v}_d^{(1)} = \frac{ e \bm{E}_{\parallel} /m^*}{1/\tau_p-i\omega} \:, \;\;\; \bm{v}_d^{(-1)} = \frac{ e \bm{E}_{\parallel}^* /m^*}{1/\tau_p+i\omega}  \:.
\end{equation}
The direct electric current is then found by multiplying $\bm{v}_d^{(0)}$ by the electron charge $e$ and the carrier density $N_e$, which yields 
\begin{equation}\label{j_dc}
\bm{j} = - N_e \frac{\zeta e^3 \tau_p^2}{m^*} \left( \frac{ \bm{E}_{\parallel} E_z^*}{1-i\omega\tau_p}   + \frac{ \bm{E}_{\parallel}^* E_z}{1+i\omega\tau_p} \right) \:.
\end{equation}
Equation~(\ref{j_dc}) contains both linear and circular currents and can be rewritten in the form of Eq.~(\ref{j_PGE_phen}) with the phenomenological parameters
\begin{equation}\label{LC}
L = - N_e \frac{\zeta e^3 \tau_p^2 /m^*}{1+(\omega \tau_p)^2} \:, \;\;\;
C = - N_e \frac{\zeta e^3 \tau_p^3 \, \omega /m^*}{1+(\omega \tau_p)^2} \:.
\end{equation}

\begin{figure}[b]
 \centering\includegraphics[width=0.38\textwidth]{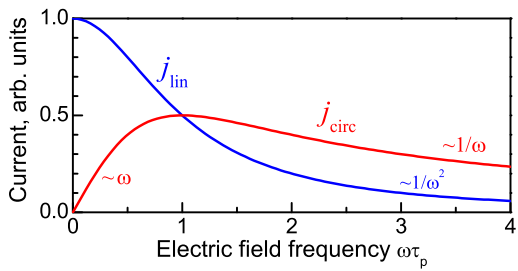}
 \caption{(Color online) Frequency dependence of electric current induced by linearly polarized and circularly polarized ac electric field.}
 \label{figure2}
\end{figure}

Shown in Fig.~\ref{figure2} are the frequency dependences of the electric currents driven by linearly polarized and circularly polarized radiation, $\bm{j}_{{\rm lin}}$ and $\bm{j}_{{\rm circ}}$, respectively. The spectral behavior of $\bm{j}_{{\rm lin}}$ repeats that of the Drude absorption: it is maximal at zero frequency and decays as $1/\omega^2$ at $\omega\tau_p \gg 1$. In contrast, the frequency dependence of the circular current is nonmonotonic. The current $\bm{j}_{{\rm circ}}$ is proportional to $\omega$ at small frequencies,
reaches maximum at $\omega\tau_p=1$ and decays as $1/\omega$ at higher frequencies.

Equations~(\ref{j_dc}) and~(\ref{LC}) accurately describe the photocurrent provided $\tau_p$ is independent of the electron energy, as it happens in the case of short-range scatterers, or the electron gas is degenerate. In the latter case, $\tau_p$ should be taken at the Fermi energy. The more general expressions for the current can be derived by solving the Boltzmann kinetic equation 
\begin{equation}\label{kineq}
\frac{\partial f(\bm{p},t)}{\partial t} + e \bm{E}_{\parallel}(t) \cdot \frac{\partial f(\bm{p},t)}{\partial \bm{p}} =  {\rm St} f(\bm{p},t) 
\end{equation}
for the electron distribution function $f(\bm{p},t)$. Here, $\bm{p}=(p_x,p_y)$ is the electron momentum and ${\rm St} f(\bm{p},t)$ is the collision integral. In the case of elastic scattering, the integral has the form
\begin{equation}\label{St}
{\rm St} f(\bm{p},t) = \sum_{\bm{p}'} \left[ W_{\bm{p}\bm{p}'} f(\bm{p}',t) - W_{\bm{p}'\bm{p}} f(\bm{p},t) \right] \:,
\end{equation}
where $W_{\bm{p}'\bm{p}}=(2\pi/\hbar) \langle |V_{\bm{p}'\bm{p}}|^2 \rangle \, \delta(\varepsilon_{\bm{p}}-\varepsilon_{\bm{p}'})$ is the rate of scattering between the states $\bm{p}$ and $\bm{p}'$, $V_{\bm{p}'\bm{p}}$ is the scattering matrix element, and the angle brackets denote averaging over the positions of impurities. 

The effect of the out-of-plane field component $E_z(t)$ on the function of size quantization 
can be treated as the field-induced admixture of excited electron states to the ground-subband wave function. To first order in the perturbation theory, the wave function of the ground subband $e1$ has the form
\begin{equation}\label{varphi}
\varphi(z,t) = \varphi_1(z) + e E_z(t) \sum_{\nu \neq 1} \frac{z_{\nu 1}}{\varepsilon_{\nu1}}  \varphi_{\nu}(z) \:,
\end{equation}
where $\varphi_{\nu}(z)$ are the wave functions along $z$ at zero electric field,
$z_{\nu 1}=\int \varphi_{\nu}(z) z \varphi_1(z) dz $ are the coordinate matrix
elements, $\varepsilon_{\nu 1}$ are energy separations between the subbands, and $\nu$ is the subband index (the dominant contribution $\propto E_z(t)$ comes usually from $\nu=2$). Then, the scattering rate can be presented in the form
\begin{equation}\label{W_total}
W_{\bm{p}'\bm{p}} = W_{\bm{p}'\bm{p}}^{(0)} + \delta W \:,
\end{equation}
where $W_{\bm{p}'\bm{p}}^{(0)} = (2\pi/\hbar) \langle | V_{11} (\bm{p}',\bm{p})|^2 \rangle \, \delta ( \varepsilon_{\bm{p}} - \varepsilon_{\bm{p}'} )$ is the scattering rate at zero electric field and $\delta W$ is the linear in $E_z(t)$ term given by
\begin{equation}\label{deltaW}
\delta W =  \frac{8 \pi e}{\hbar} E_z(t) \sum_{\nu \neq 1} \frac{z_{\nu 1}}{\varepsilon_{\nu 1}} \langle {\rm Re} \, V_{11}^* V_{1\nu} \rangle \, \delta(\varepsilon_{\bm{p}}-\varepsilon_{\bm{p}'}) \:,
\end{equation}
with $V_{11}$ and $V_{1\nu}$ being the ``intrasubband'' and ``intersubband'' matrix elements of scattering. The rate $W_{\bm{p}'\bm{p}}^{(0)}$ is determined by intrasubband scattering and may depend on the initial $\bm{p}$ and final $\bm{p}'$ electron momenta. In contrast, the scattering processes between states described by different functions $\varphi_{\nu}(z)$ require the transfer of momentum comparable to $\pi \hbar/a$ ($a$ is the quantum well width) which is much larger than the in-plane electron momentum. Such processes can be caused by short-range scatterers only, therefore, we assume that $\delta W$ is independent of the directions of $\bm{p}$ and $\bm{p}'$. In this case, the contribution to the collision integral~(\ref{St}) proportional to $E_z(t)$ has the form
\begin{equation}\label{delta_St}
\delta {\rm St} f(\bm{p},t) = - \zeta e E_z(t) [f(\bm{p},t) - \bar{f}(\bm{p},t)] \:,
\end{equation}
where $\bar{f}(\bm{p},t)$ is the distribution function averaged over the directions of $\bm{p}$ and 
\begin{equation}\label{zeta}
\zeta = \frac{4 m^*}{\hbar^3}  \sum_{\nu \neq 1} \frac{z_{\nu 1}}{\varepsilon_{\nu 1}}  \langle {\rm Re} \, V_{11}^* V_{1\nu} \rangle \:.
\end{equation}
We note that the products $z_{\nu 1} \langle {\rm Re} \, V_{11}^* V_{1\nu} \rangle$ and, hence, the parameter $\zeta$ are equal to zero in absolutely symmetric structures where the impurity profile is an even function with respect to the QW center and the functions $\varphi_{\nu}$ are either odd or even.

To solve the kinetic equation~(\ref{kineq}) with the scattering rate~(\ref{W_total})
we decompose the distribution function $f(\bm{p},t)$ into frequency $n$ and angular $m$ harmonics as follows
\begin{equation}\label{f_harmonics}
f(\bm{p},t) = \sum\limits_{n,m} f_{m}^{(n)} \exp( i m \theta_{\bm{p}} - i n \omega t) \:,
\end{equation}
where $\theta_{\bm{p}}$ is the polar angle of the vector $\bm{p}$.  Then, for the time-independent asymmetric part of the distribution function $\delta f(\bm{p})= f_{1}^{(0)} \exp( i \theta_{\bm{p}}) + f_{-1}^{(0)} \exp( -i \theta_{\bm{p}})$, which determines dc current, we derive
\begin{equation}\label{f_harmonics2}
\delta f(\bm{p}) = \zeta e^2 \tau_p^2 \left[ \frac{(\bm{E}_{\parallel} \cdot \bm{v}) E_z^*}{1-i\omega\tau_p} +\frac{(\bm{E}_{\parallel}^* \cdot \bm{v}) E_z}{1+i\omega\tau_p} \right] \frac{d f_{\varepsilon}}{d \varepsilon} \:,
\end{equation}
where $f_{\varepsilon}$ is the equilibrium distribution function of carriers, $\varepsilon=\bm{p}^2/(2m^*)$ and $\bm{v}=\bm{p}/m^*$ are is the electron kinetic energy and velocity, respectively, the momentum relaxation time is given by $\tau_p^{-1} = \sum_{\bm{p}'} W_{\bm{p}'\bm{p}}^{(0)} (1-\cos\theta_{\bm{p}'\bm{p}})$, and
$\theta_{\bm{p}'\bm{p}}=\theta_{\bm{p}'}-\theta_{\bm{p}}$ is the angle between $\bm{p}'$ and $\bm{p}$.

The electric current is obtained from Eq.~(\ref{f_harmonics2}) by multiplying $\delta f(\bm{p})$ by the electron charge $e$ and velocity $\bm{v}$, and summing up the result over the momenta $\bm{p}$. It gives 
\begin{equation}\label{j_final}
\bm{j} = \frac{2 \zeta e^3}{m^*} \sum_{\bm{p}} \tau_p^2 \left( \frac{ \bm{E}_{\parallel} E_z^*}{1-i\omega\tau_p} +\frac{\bm{E}_{\parallel}^* E_z }{1+i\omega\tau_p} \right) \frac{\varepsilon \, d f_{\varepsilon}}{d \varepsilon}  \:,
\end{equation}
where the factor of 2 accounts for the spin degeneracy. Equation~(\ref{j_final}) accurately takes into account the possible dependence of the momentum relaxation time on energy and is more general than Eq.~(\ref{j_dc}). In the case of short-range scattering, where energy dependence of $\tau_p$ vanishes, Eqs.~(\ref{j_final}) and~(\ref{j_dc}) are equivalent to each other and can also be rewritten in the form of Eq.~(\ref{j_PGE_phen}) with the parameters  
\begin{equation}\label{L_final}
L = - \frac{4 N_e e^3 \tau_p /m^*}{1+(\omega\tau_p)^2}  \sum_{\nu \neq 1} \frac{z_{\nu 1}}{\varepsilon_{\nu 1}}  \xi_{\nu} \:, 
\end{equation}
\begin{equation}\label{C_final}
C = - \frac{4 N_e e^3 \tau_p^2 \,\omega /m^*}{1+(\omega\tau_p)^2} \sum_{\nu \neq 1} \frac{z_{\nu 1}}{\varepsilon_{\nu 1}}  \xi_{\nu} \:, 
\end{equation}
where
\begin{equation}\label{xi}
\xi_{\nu} = \frac{\langle {\rm Re} \, V_{11}^* V_{1\nu} \rangle}{\langle |V_{11}|^2 \rangle} = \frac{\int_{-\infty}^{\infty} \varphi_1^3(z) \varphi_{\nu}(z) u(z) dz }{\int_{-\infty}^{\infty} \varphi_1^4(z) u(z) dz} \:,
\end{equation}
with $u(z)$ being the profile of impurity distribution along the QW growth direction. Following Eqs.~(\ref{j_PGE_phen}), (\ref{L_final}), and~(\ref{C_final}) one can estimate the electric current magnitude. The estimate gives $j \sim 10^{-3}$~A/cm for a GaAs-based quantum well with the electron density $N_e=10^{11}$~cm$^{-2}$, the well width $20$~nm, the structure asymmetry degree $\xi_2=0.1$, the electric field amplitude $E_z=E_{\parallel}=1$~kV/cm, and the frequency $\omega = 1/\tau_p$.

The above developed microscopic theory describes the formation of dc current by ac electric field in the whole range of classical frequencies, no matter how large is $\omega\tau_p$. At even higher frequencies, when $\hbar\omega$ becomes comparable to the mean electron energy $\bar{\varepsilon}$, another approach involved quantum optical transitions is required. However, both classical and quantum approaches unite and should give the same results in the intermediate frequency range $1/\tau_p \ll \omega \ll \bar{\varepsilon}/\hbar$. This is indeed the case: our results for the circular current [Eq.~(\ref{C_final})] in the limit $\omega\tau_p \gg 1$ coincide with those obtained in Ref.~\cite{Tarasenko07}, 
where virtual indirect optical transitions were considered. 
The linear current [Eq.~(\ref{L_final})] decreases as $1/(\omega^2 \tau_p)$ at $\omega\tau_p \gg 1$ and becomes much smaller than the circular current at $\hbar\omega \approx \bar{\varepsilon}$. The microscopic theory of the linear photocurrent in this spectral range is a task for future. 

It is also worth mentioning that, in QWs grown from zinc-blende-type semiconductors, there are additional contributions to linear and circular currents caused by bulk inversion asymmetry.
Those currents flow in different directions with respect to the currents caused by structure inversion asymmetry and can be easily discriminated in experiment~\cite{Giglberger07}. The presented model of high-frequency nonlinear conductivity can be applied to evaluate those currents as well. In this case,
one should consider the modulation of electron mobility caused by the electric-field-induced admixture of the valence-band states to the electron wave function~\cite{Tarasenko07}. Finally, we note that the effect of electric field on the mobility can originate not only from the change of scattering rate but also from the variation of the effective electron mass as was proposed in Ref.~\cite{Takhtamirov09}.

\section{Magnetic-field-induced currents}\label{MPGE}

The application of a static magnetic field $\bm{B}$ in the QW plane enables the generation of a direct current even in the geometry where ac electric field oscillates in the interface plane. Within linear in $\bm{B}$ approximation, such currents are described by the second term of the right-hand side of phenomenological Eq.~(\ref{j_phen}). Symmetry analysis shows~\cite{IvchenkoGanichev,Belkov05} that, for this particular geometry, the polarization dependence of the electric current caused by QW structure inversion asymmetry is given by
\begin{eqnarray}\label{j_MPGE}
j_x &=& M_1 [B_y (|E_x|^2-|E_y|^2) - B_x (E_x E_y^* + E_y E_x^* ) ] \nonumber \\
&+& M_2 B_y |\bm{E}|^2 +  M_3 B_x i (E_x E_y^* - E_y E_x^*)  \:, \\
j_y &=& M_1 [B_x (|E_x|^2-|E_y|^2) + B_y (E_x E_y^* + E_y E_x^* ) ] \nonumber \\
&-& M_2 B_x |\bm{E}|^2 + M_3 B_y i (E_x E_y^* - E_y E_x^*) \:. \nonumber
\end{eqnarray} 
Here, the parameter $M_1$ describes the electric current whose magnitude and direction depend on linear polarization of the field, $M_2$ describes the polarization-independent current, and $M_3$ is responsible for the circular current sensitive to the radiation helicity.

Microscopically, magnetic-field-induced currents can be of both diamagnetic and paramagnetic (spin-depen\-dent) origins. Spin-dependent mechanisms of the current generation are based on the Zeeman splitting of electron states in the magnetic field together with spin-dependent electron scattering; they are studied for $\omega\tau_p \gg 1$ in Refs.~\cite{Belkov05,Ganichev06,Ganichev09}. 
Below we focus on diamagnetic mechanisms, which are less investigated and 
do not require spin-orbit coupling.

\begin{figure}[t]
 \centering\includegraphics[width=0.43\textwidth]{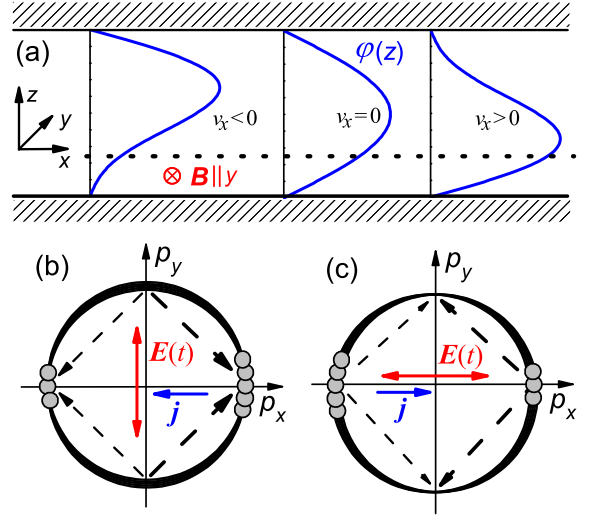}
 \caption{(Color online) (a) Asymmetry in electron scattering by impurities caused by in-plane magnetic field. (b) and (c) microscopic model of dc current generation due to electric-field-induced alignment of electron momenta followed by asymmetric scattering.}
 \label{figure3}
\end{figure}

The diamagnetic mechanisms of the current generation are based on the asymmetry in electron scattering by static defects or phonons in quantum wells subjected to an in-plane magnetic field. Such a scattering asymmetry in $\bm{p}$-space originates from the Lorentz force acting upon mobile carriers and modifying their wave functions~\cite{Tarasenko08}, which is illustrated in Fig.~\ref{figure3}a.
Here, we assume that electrons are scattered by impurities, the structure inversion asymmetry of QW is modeled by placing the $\delta$-layer of impurities (dotted line) closer to the lower interface, and the magnetic field $\bm{B} \parallel y$. Electrons with different velocities move in the QW plane and are pushed by the Lorentz force $\bm{F}_L = (e/c)[\bm{v}\times\bm{B}]$ to the lower or upper interface depending on the sign of $v_x$. This leads to a variation of the electron function of size quantization which results, in turn, in the asymmetry of electron scattering: carriers with $v_x>0$ are scattered by impurities at higher rate than those with $v_x < 0$. The Lorentz force is proportional to both the magnetic field and the electron velocity, therefore, the small correction to the
scattering rate is linear in $\bm{p}$ and $\bm{B}$. In quantum wells with structure inversion asymmetry, the rate of elastic electron scattering can be presented in the form~\cite{Tarasenko08}
\begin{equation}\label{W_scat_B}
W_{\bm{p}'\bm{p}} = W_{\bm{p}'\bm{p}}^{(0)}  + w'[B_x (p_y + p'_y) - B_y (p_x + p'_x)] \:,
\end{equation}
where 
\begin{equation}\label{w_prime}
w' = - \frac{4 \pi e}{\hbar m^* c} \sum_{\nu \neq 1} \frac{z_{\nu 1}}{\varepsilon_{\nu 1}} \langle {\rm Re} \, V_{11}^* V_{1\nu} \rangle \, \delta(\varepsilon_{\bm{p}}-\varepsilon_{\bm{p}'}) \:.
\end{equation}
The parameter $w'$ is determined by the same matrix elements of scattering as $\zeta$, see Eq.~(\ref{zeta}). Below it is assumed that $w'$ is independent of the directions of $\bm{p}$ and $\bm{p}'$. 

The asymmetry in electron scattering gives rise to a direct current if the carriers are excited by ac electric field. Figures~\ref{figure3}b and~\ref{figure3}c illustrate a mechanism of current formation for linearly polarized electric field oscillating along the $y$ and $x$ axes, respectively.
The ac electric field leads, in the second order in the field amplitude $\bm{E}$, to the alignment of electron momenta along the axis of field oscillations. In the case of $\bm{E} \parallel y$, the carriers populate predominantly the states with large $|p_y|$, 
which is indicated in Fig.~\ref{figure3}b by the Fermi circle of variable thickness. The processes of electron scattering shown by dashed lines suppress the alignment tending to restore the isotropic distribution of carriers in $\bm{p}$-space. However, in the presence of magnetic field $B_y$, the rates of electron scattering 
to the states with positive and negative $p_x$ are different, see Eq.~(\ref{W_scat_B}). Such a difference in scattering rates is illustrated in Fig.~\ref{figure3}b by lines of different thicknesses. Therefore, the scattering events result in an imbalance of carrier population between positive and negative $p_x$ (shown by full circles) giving rise to an electric current $j_x$.

Figure~\ref{figure3}c sketches the same mechanism of the current formation for $\bm{E} \parallel x$. In this case, ac electric field leads to the alignment of electron momenta along the $x$ axis. The subsequent asymmetric scattering of electrons in the magnetic field $B_y$ also causes imbalance in the carrier population between positive and negative $p_x$ giving rise to a dc current $j_x$. However, as it follows from Figs.~\ref{figure3}b and~\ref{figure3}c, the current directions are opposite for $\bm{E} \parallel y$ and $\bm{E} \parallel x$. Thus, the illustrated mechanism of the current formation describes the polarization-dependent contribution to the current which is given by the term $\propto M_1$ in Eq.~(\ref{j_MPGE}).

The analytical expression for magnetic-field-induced currents can be derived
in the framework of kinetic theory by solving Eq.~(\ref{kineq}) with the scattering rate (\ref{W_scat_B}). To solve the equation we decompose the distribution function $f(\bm{p},t)$ into frequency $n$ and angular $m$ harmonics according to Eq.~(\ref{f_harmonics}) and obtain the system of coupled equations. Its solution for the time-independent asymmetric part of the distribution function $\delta f(\bm{p})$, which determines dc current, in the geometry $\bm{B} \parallel y$ assumes the form
\begin{eqnarray}\label{delta_f_B}
\delta f (\bm{p}) = &-& \left[ (|E_x|^2+|E_y|^2) v_x - (E_x E_y^* - E_y E_x^*) v_y \right] \nonumber \\
&\times& \frac{\zeta \tau_p \, e^3 B_y}{4 m^* c} \left( 2 + \varepsilon \frac{d}{d \varepsilon} \right) \frac{\tau_p \tau_ 2 \, d f_{\varepsilon} / d\varepsilon}{(1-i\omega\tau_p)(1-i\omega\tau_2)} \nonumber \\
&-& \left[ (|E_x|^2-|E_y|^2) v_x + (E_x E_y^* + E_y E_x^*) v_y \right] \nonumber \\
&\times& \frac{\zeta \tau_p \tau_2 \, e^3 B_y}{4 m^* c} \left( \varepsilon  \frac{d}{d \varepsilon} \right) \frac{\tau_p \, d f_{\varepsilon} /d\varepsilon}{1-i\omega\tau_p} + {\rm c.c.} \:,
\end{eqnarray}
where $\tau_2$ is the relaxation time of the second angular harmonic of the electron distribution function, $\tau_2^{-1} = \sum_{\bm{p}'} W_{\bm{p}'\bm{p}}^{(0)} (1-\cos2\theta_{\bm{p}'\bm{p}})$. 

The current density $\bm{j}$ is obtained by multiplying $\delta f(\bm{p})$ by the electron charge and velocity, and summing up the result over the momentum. This procedure yields Eq.~(\ref{j_MPGE}) with the following parameters
\begin{equation}\label{M1}
M_1 = \frac{\zeta e^4}{c \, m^{*2}} \sum_{\bm{p}} \frac{ \tau_p \, d (\tau_p \tau_2 \, \varepsilon^2)/d \varepsilon }{1+(\omega \tau_p)^2} \frac{d f_{\varepsilon}}{d \varepsilon} \:, 
\end{equation}
\vspace{-0.5cm}
\begin{equation}\label{M2}
M_2 = \frac{\zeta e^4}{c \, m^{*2}} \sum_{\bm{p}} \frac{(1-\omega^2 \tau_p \tau_2) \,  \tau_p \tau_2 \, \varepsilon^2 \tau'_p }{[1+(\omega \tau_p)^2][1+(\omega \tau_2)^2]} \frac{d f_{\varepsilon}}{d \varepsilon} \:,
\end{equation}
\vspace{-0.5cm}
\begin{equation}\label{M3}
M_3 = - \frac{\zeta e^4}{c \, m^{*2}} \sum_{\bm{p}} \frac{ \omega \tau_p \tau_2 (\tau_p+\tau_2) \, \varepsilon^2 \tau'_p }{[1+(\omega \tau_p)^2][1+(\omega \tau_2)^2]} \frac{d f_{\varepsilon}}{d \varepsilon} \:,
\end{equation}
where $\tau'_p = d \tau_p / d \varepsilon$, and the factor 2 of spin degeneracy is already taken into account. It follows from Eqs.~(\ref{j_MPGE}) and~(\ref{M1})-(\ref{M3}) that the polarization-independent and circular currents given by $M_2$ and $M_3$, respectively, emerge due to energy dependence of the momentum relaxation time. If the energy dependence of $\tau_p$ and $\tau_2$ can be neglected and $\tau_p=\tau_2$,
the parameter $M_1$ assumes the form
\begin{equation}\label{M_1}
M_1 = - 4 \frac{N_e e^4 \tau_p^2 /(c m^{*2})}{1+(\omega\tau_p)^2} \sum_{\nu \neq 1} \frac{z_{\nu 1}}{\varepsilon_{\nu 1}}  \xi_{\nu} \:, 
\end{equation}
while $M_2$ and $M_3$ vanish. We also note that there are additional contributions to the polarization-independent current which originate from the energy relaxation of hot carriers in magnetic field~\cite{Tarasenko08}. These contributions depend on the energy relaxation mechanisms and are out of scope of the present paper.

The magnetic-field-induced current~(\ref{j_MPGE}) can be considerably larger than the current~(\ref{j_PGE_phen}) excited at oblique polarization of ac electric field at $B=0$. For linearly polarized radiation, the ratio between the current magnitudes is estimated as $M_1 B / L = e B \tau_p /(m^* c)$, which can exceed unity already at moderate magnetic fields.

In conclusion, we have developed the microscopic theory of high-frequency nonlinear conductivity in doped quantum wells with structure inversion asymmetry. It is shown that the excitation of carriers by ac electric field in such structures leads to a dc current, its direction and magnitude being determined by the field polarization. The theory can also be applied to study the nonlinear generation of combination frequencies (frequency mixing). 

\paragraph*{Acknowledgments.}

The author acknowledges useful discussions with E.L.~Ivchenko and L.E.~Golub. 
This work was supported by the RFBR, the President Grant for young scientists (MD-1717.2009.2), and the Foundation ``Dynasty''-ICFPM.


\begin{thebibliography}{99}

\bibitem{Sturman_book} B.I.~Sturman and V.M.~Fridkin, 
\textit{The Photovoltaic and Photorefractive Effects in
Non-Centrosymmetric Materials} (Gordon and Breach Science
Publishers, New York, 1992).
 
\bibitem{IvchenkoGanichev} E.L. Ivchenko and S.D.~Ganichev, 
\textit{Spin Photogalvanics} in \textit{Spin Physics in Semiconductors}, 
ed. M.I.~Dyakonov, (Springer, Berlin, 2008). 

\bibitem{Giglberger07} S.~Giglberger, L.E.~Golub, V.V.~Bel'kov, S.N.~Danilov,
D.~Schuh, C.~Gerl, F.~Rohlfing, J.~Stahl, W.~Wegscheider, D.~Weiss, W.~Prettl,
and S.D.~Ganichev, 
Phys. Rev. B {\bf 75}, 035327 (2007).

\bibitem{Olbrich09} P.~Olbrich, J.~Allerdings, V.V.~Bel'kov, S.A.~Tarasenko, D.~Schuh, W.~Wegscheider, T.~Korn, C.~Sch\"{u}ller, D.~Weiss, and S.D. Ganichev,
Phys. Rev. B {\bf 79}, 245329 (2009).

\bibitem{Zhang09} Q.~Zhang, X.Q.~Wang, X.W.~He, C.M.~Yin, F.J.~Xu, B.~Shen, Y.H.~Chen, Z.G.~Wang, Y.~Ishitani, and A.~Yoshikawa, 
Appl. Phys. Lett. {\bf 95}, 031902 (2009). 

\bibitem{Dai10} J.~Dai, H.-Zh.~Lu, C.L.~Yang, Sh.-Q.~Shen, F.-Ch.~Zhang, and X.~Cui,
Phys. Rev. Lett. {\bf 104}, 246601 (2010).

\bibitem{Tarasenko07} S.A.~Tarasenko,
Pis'ma v ZhETF {\bf 85}, 216 (2007) [JETP Lett. {\bf 85}, 182 (2007)]. 

\bibitem{Takhtamirov09} E.~Takhtamirov,
arXiv:0904.2341.

\bibitem{Reimann02} P.~Reimann,
Phys. Rep. {\bf 361}, 57 (2002).

\bibitem{Entin06} M.V.~Entin and L.I.~Magarill,
Phys. Rev. B {\bf 73}, 205206 (2006).

\bibitem{Sassine08} S.~Sassine, Yu.~Krupko, J.-C.~Portal, Z.D.~Kvon, R.~Murali, K.P.~Martin,
G.~Hill, and A.D.~Wieck, 
Phys. Rev. B {\bf 78}, 045431 (2008).

\bibitem{Deyo09} E.~Deyo, L.E.~Golub, E.L.~Ivchenko, and B.~Spivak,
arXiv:0905.3295.

\bibitem{Moore10} J.E.~Moore and J.~Orenstein,
Phys. Rev. Lett. {\bf 105}, 026805 (2010). 

\bibitem{Belkov05} V.V.~Bel'kov, S.D.~Ganichev,  E.L.~Ivchenko, S.A.~Tarasenko,  W.~Weber,
S.~Giglberger,  M.~Olteanu, P.~Tranitz, S.N.~Danilov, P.~Schneider, W.~Wegscheider,
D.~Weiss, and W.~Prettl,
J. Phys.: Condens. Matter {\bf 17}, 3405 (2005). 

\bibitem{Ganichev06} S.D.~Ganichev, V.V.~Bel'kov, S.A.~Tarasenko, S.N.~Danilov, S.~Giglberger, Ch.~Hoffmann, E.L.~Ivchenko, D.~Weiss, W.~Wegscheider, C.~Gerl, D.~Schuh, J.~Stahl, J.~De~Boeck, G.~Borghs, and W.~Prettl,
Nature Phys. {\bf 2}, 609 (2006).

\bibitem{Ganichev09} S.D.~Ganichev, S.A.~Tarasenko, V.V.~Bel'kov, P.~Olbrich, W.~Eder, D.R.~Yakovlev, V.~Kolkovsky, W.~Zaleszczyk, G.~Karczewski, T.~Wojtowicz, and D.~Weiss,
Phys. Rev. Lett. {\bf 102}, 156602 (2009).

\bibitem{Tarasenko08}  S.A.~Tarasenko,
Phys. Rev. B {\bf 77}, 085328 (2008).

\end{thebibliography}
\end{document}